\documentclass{elsart}
\usepackage{graphicx}
\begin{document}
\begin{frontmatter}
\journal{Surface Science Letters}
\title{A spin-polarised first principles study of short dangling bond wires
on Si(001)}
\author{C.F.~Bird} and 
\ead{c.bird@ucl.ac.uk}
\ead[url]{http://www.cmmp.ucl.ac.uk/$\sim$cfb/}
\author{D.R.~Bowler\corauthref{drb}\thanksref{LCN}}
\ead{david.bowler@ucl.ac.uk}
\ead[url]{http://www.cmmp.ucl.ac.uk/$\sim$drb/research.html}
\corauth[drb]{Corresponding author}
\address{Department of Physics and Astronomy, University College London,
Gower Street, London WC1E 6BT, UK}
\thanks[LCN]{Also at: London Centre for Nanotechnology, Department of Physics and Astronomy, Gower Street, 
London WC1E 6BT, UK}
\begin{abstract}
Short dangling bond wires (DB wires), fabricated on H-terminated
Si(001) surfaces, show patterns of displacement that depend on their
length.  We have performed density function calculations, with and
without spin-polarisation, designed to investigate the atomic and
electronic structure of these wires.  
As expected, we find that even length wires are accurately modelled
by non-spin polarised calculations, whilst odd length wires must be modelled
using spin-polarised calculations. In particular, the use of spin-polarisation
provides quantative agreement with STM observations, rather than
the qualitative agreement reported elsewhere. 
\end{abstract}

\end{frontmatter}

\section{Introduction}

\begin{figure}
\includegraphics[width=\textwidth]{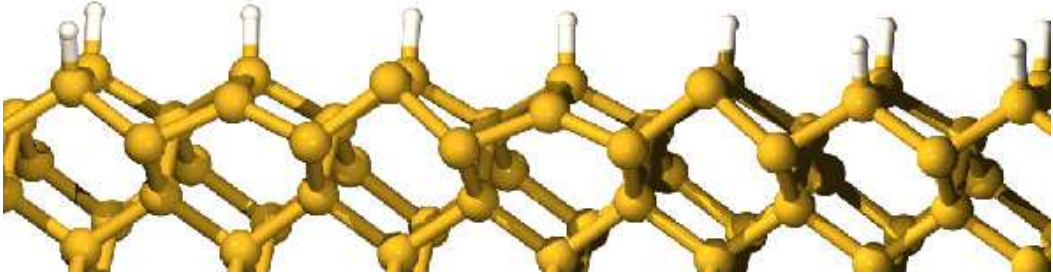}
  \caption{\label{fig:Len4}A DB wire of length four, showing the DUDU
  pattern with one terminating hydrogen on the left and two on the
  right.  Only the top four layers of silicon are shown.}
\end{figure}

With the continuing drive to further miniaturize microelectronics and
develop realistic nanoelectronic devices, there is a great deal of
interest in the atomic and electronic properties of atomic scale
devices and the change in behaviour at this scale.  One prototypical
model for an atomic scale wire is the dangling bond 
wire\cite{Shen1995}, which is formed by removing a line of hydrogen
atoms from one side of a dimer row on a hydrogenated Si(001) surface
with a STM tip.  The atomic structure of a length four DB wire is shown in
Fig.~\ref{fig:Len4}.  We present electronic structure calculations designed to investigate the atomic
and electronic structure of short, finite dangling bond wires, and show
that for these wires there is a marked difference between wires with
odd and even numbers of atoms.

The infinite dangling bond wire undergoes a Peierls
transition\cite{Hitosugi1999,peierls1955,Bowler2001}, with alternate atoms
displacing up and down together with charge transferring from the down atoms to
the up atoms.  Finite wires should undergo a Jahn-Teller distortion\cite{Jahn1937} 
(since Peierls distortions only apply to infinite one-dimensional 
systems),
and there is good STM evidence that this occurs\cite{Hitosugi1999},
although the exact mechanism of the distortion has recently been
contested\cite{Cho2002}.
Conduction in these one-dimensional systems is expected to involve
polaronic effects, and it has been shown theoretically that injection
of charge into the dangling bond wire results in formation of
polarons\cite{Bowler2001}, which are highly mobile despite their
localisation\cite{todorovic02}.  We also expect soliton effects, which we 
have investigated elsewhere\cite{Bird2003}.

We have performed calculations within density functional theory (DFT) using
spin polarisation in the generalized-gradient approximation (GGA), 
implemented with pseudopotentials and a plane-wave basis set.  We have 
calculated the atomic and electronic structure of wires from two to five
atoms long.  It is important to note that the odd-length wires have an
unpaired electron, and thus the use of spin polarisation is vitally 
important (and has a large effect on the results, as discussed below).

There has been previous modelling work on this
system\cite{Hitosugi1999,Bowler2001,Cho2002,Watanabe1997}, but none of this work has considered
the importance or effect of spin polarisation.  Bowler and
Fisher\cite{Bowler2001} and Cho \& Kleinman\cite{Cho2002}
considered flat and infinite wires with LDA and GGA respectively, and found the relative
displacement of up and down atoms to be 0.67\,\AA and 0.74\,\AA
respectively.  Watanabe \textit{et
al.\/}\cite{Watanabe1997} considered infinite DB wires of different
types (and their interaction with Ga), while Hitosugi \textit{et
al.\/}\cite{Hitosugi1999} considered the same finite wires that we
model in this paper. They compared the simulation results to STM
images of the wires, finding qualitative agreement: the length 3 DB
wire shows corrugation of $\sim0.15$\,\AA\ in STM, while their
calculations find a corrugation of $\sim0.50$\,\AA. 
We will show below that agreement between simulations and observations is 
improved enormously by the inclusion of spin polarisation in the calculation.

The rest of the paper is organised as follows: in the next section, we 
present details of the computational procedure that we used; we then present
the results of our modelling for the DB wires; and finally we present our
conclusions.

\section{Computational Details}
We performed DFT calculations using the PW91 GGA functional\cite{pw91} for
exchange and correlation, both with and without spin-polarisation.  We
used the VASP code\cite{Kresse1996}, with ultrasoft pseudopotentials,
a plane wave cutoff of 200 eV and a 2$\times$2$\times$1
Monkhorst-Pack\cite{monkhorst76a} grid of \textbf{k}-points; these parameters give good
energy and force convergence.  The unit cell was 8 dimers long, one
dimer row wide and six layers deep, with the bottom layer of atoms
constrained to remain in bulk-like positions and terminated with
hydrogen, giving a total of 136-144 atoms (going from an infinite wire
to a totally hydrogenated surface). Calculations on even length wires
(including the 8 atom cell used for the infinite wire)
using spin-polarisation were constrained to a total magnetic moment of
zero.  While odd length wires should be only correctly be modelled
with spin polarisation, we performed non-spin polarised calculations
for comparison purposes (similarly, the spin polarised calculations of
even length wires were run for comparison).

\section{Results}

\begin{table}
\caption{\label{tab:Energies}Minimum total energies for different
length dangling bond wires with and without spin polarisation}
\begin{tabular}{l|c|c|c|c|c|c}
Spin ? & Length 2 & Length 3 & Length 4 & Length 5 & Infinite & Hydrogenated\\
\hline
No     & -664.742 & -660.211 & -655.776 & -651.235 & -638.112 & -673.876\\
Yes    & -664.749 & -660.308 & -655.694 & -651.297 & -637.691 & -673.876\\
\hline
\end{tabular}
\end{table}

The energetics for DB wires of lengths 2, 3, 4, 5 and an infinite
wire, found both with and without spin polarisation, are presented in
Table~\ref{tab:Energies}.  This shows that the lowest energy
structures (which will be analysed in detail below) for even
length wires are found without spin polarisation and for odd length 
wires are found with spin polarisation, as might be expected.

\begin{table}
\caption{\label{tab:SpinDisplacements}Vertical displacements of
atoms in dangling bond wires of different length, found using spin
polarisation.  Displacements are given in \AA\ relative to hydrogenated 
atoms. Atoms 5-8 apply only to the atoms in the `infinite' 8 atom cell.}
\begin{tabular}{l|c|c|c|c|c|c|c|c}
Length & Atom 1 & 2 & 3 & 4 & 5 & 6 & 7 & 8 \\
\hline
2      & -0.080 & -0.049 &    --- &    --- &    --- & -- & -- & -- \\
3      & -0.016 & -0.148 & -0.016 &    --- &    --- & -- & -- & -- \\
4      &  0.007 & -0.143 &  0.111 & -0.013 &    --- & -- & -- & -- \\
5      &  0.000 & -0.132 &  0.000 & -0.132 & -0.016 & -- & -- & -- \\
Inf    & -0.058 & -0.063 & -0.083 & -0.020 & -0.113 & -0.036 & -0.068 & -0.060 \\
\hline
\end{tabular}
\end{table}

The displacement patterns of atoms in these wires are given in
Table~\ref{tab:SpinDisplacements} for spin polarised calculations, and
Table~\ref{tab:NonSpinDisplacements} for non-spin polarised
calculations.  The displacements are significantly larger without spin
polarisation than with spin polarisation, and this agrees with
experimental results.  STM measurements of short and long
wires\cite{Hitosugi1999} find a relative displacement from up to down
atoms of $\sim 0.6$\,\AA\ in a 13 atom long wire, and 0.18\,\AA\ in
short wires.  When considering STM measurements, particularly of
semiconductor surfaces, one has to take into account the fact that
there may well be contributions to measured heights from electronic as
well as geometric effects.  In the present situation, there are two
factors that give us confidence that the height measured by STM
linescans is the true height: first, the energies of the top-most
filled states for both up and down spins, which contribute strongly to
the observed features, differ in energy by only 0.01\,eV (i.e. a
negligible amount); second, the images are taken at a large bias
voltage (-2\,V) where any remaining electronic effects would be
largely washed out in STM line scans\cite{Hofer2001}.  With the
exception of the length four wire (which will be discussed below), the
simulated short wire displacements in
Table~\ref{tab:SpinDisplacements} for length three and five wires
match extremely well with the measured displacements.  The previous
modelling of these systems\cite{Hitosugi1999} found displacements of
$\sim 0.5$\,\AA, which are a factor of four larger than the measured
values, and resemble non-spin polarised calculations.  We also note
that our infinite wire without spin polarisation matches the long wire
displacements very well.

\begin{table}
\caption{\label{tab:NonSpinDisplacements}Vertical displacements of
atoms in dangling bond wires of different length, found without spin
polarisation.  Displacements are given in \AA\ relative to hydrogenated 
atoms.}
\begin{tabular}{l|c|c|c|c|c|c|c|c}
Length & Atom 1 & 2 & 3 & 4 & 5 & 6 & 7 & 8\\
\hline
2      & -0.057 & -0.036 &    --- &    --- &    --- & -- & -- & --\\
3      &  0.080 & -0.376 &  0.096 &    --- &    --- & -- & -- & --\\
4      &  0.243 & -0.441 &  0.226 & -0.344 &    --- & -- & -- & --\\
5      &  0.131 & -0.386 &  0.212 & -0.375 &  0.114 & -- & -- & --\\
Inf    & -0.479 &  0.310 & -0.481 &  0.310 & -0.477 &  0.310 & -0.481 &  0.310 \\
\hline
\end{tabular}
\end{table}

The length four wire leaves some questions to be answered.  The lowest
energy is found without spin polarisation, as expected, but this has
displacements from atom to atom of $0.6$\,\AA, compared to experimental
values of $0.1$\,\AA.  However, a length four wire will have two
different forms, DUDU and UDUD, as well as soliton-like excitations
such as UDDU (which we confirmed as existing, and being only 0.07\,eV
less stable than the ground state) which suggests that the STM images
are likely to be measuring an average of all these states, leading to
the apparent discrepancy.  A similar mechanism has been proposed
before\cite{Hitosugi1999}, though without the UDDU calculation. 

\begin{figure}
\includegraphics[angle=-90, width=0.49\columnwidth, clip]{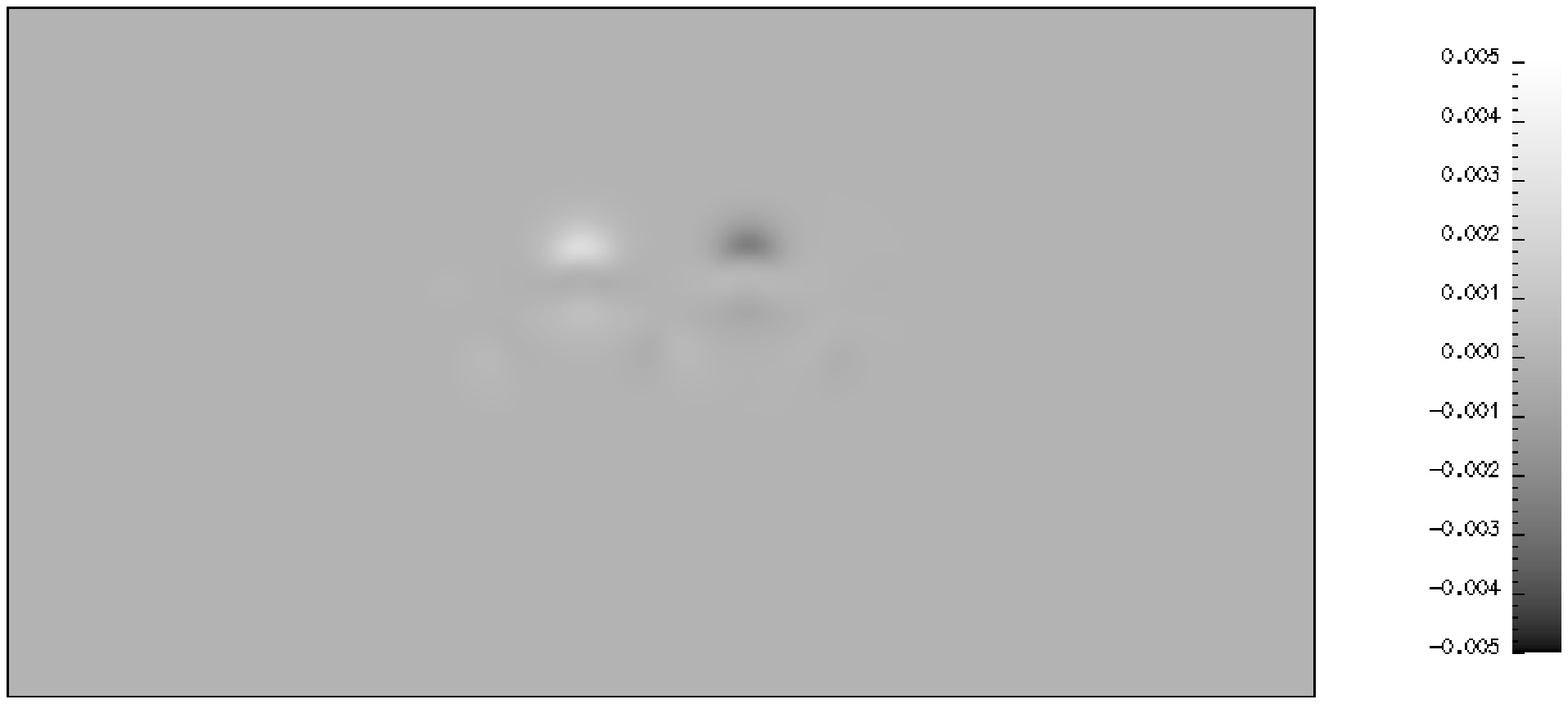}
\includegraphics[angle=-90, width=0.49\columnwidth, clip]{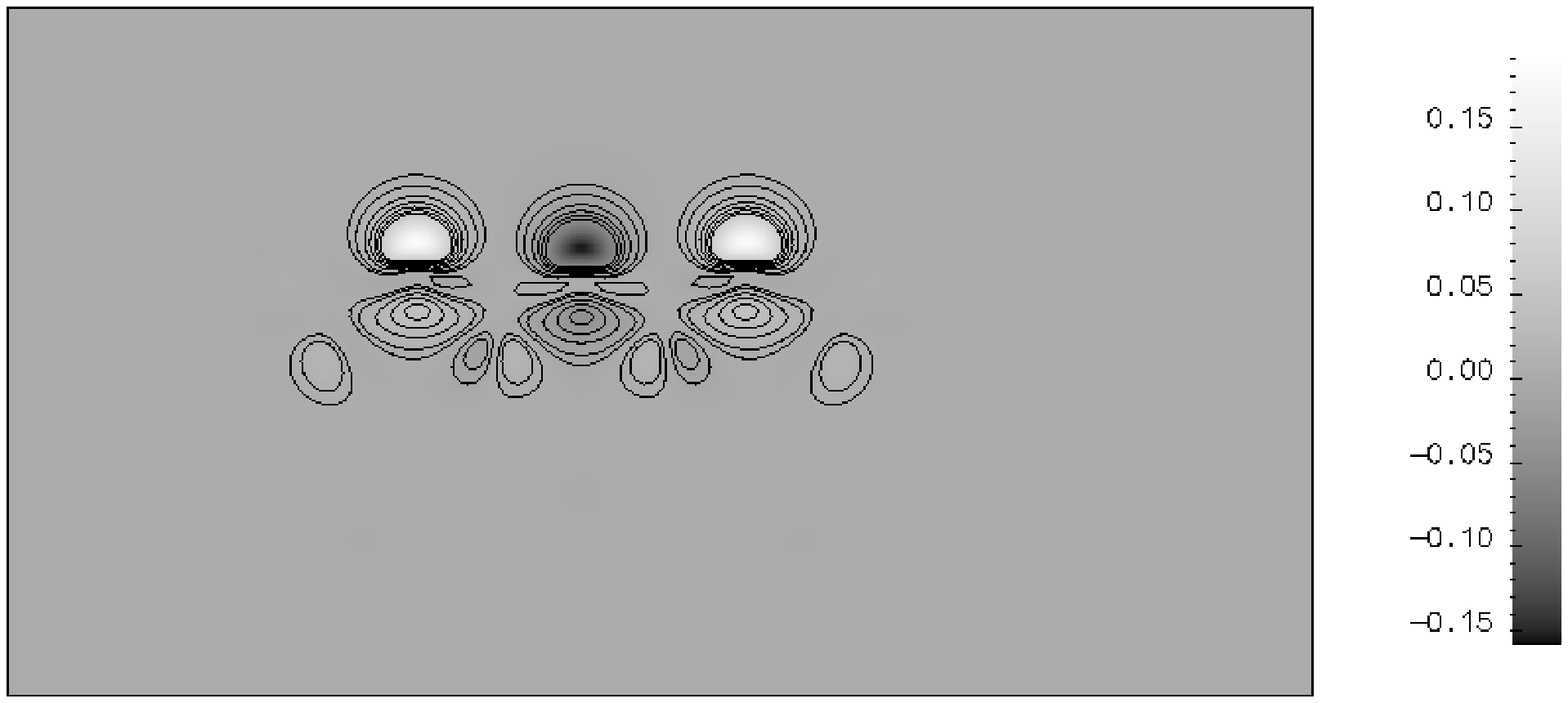}\\
\includegraphics[angle=-90, width=0.49\columnwidth, clip]{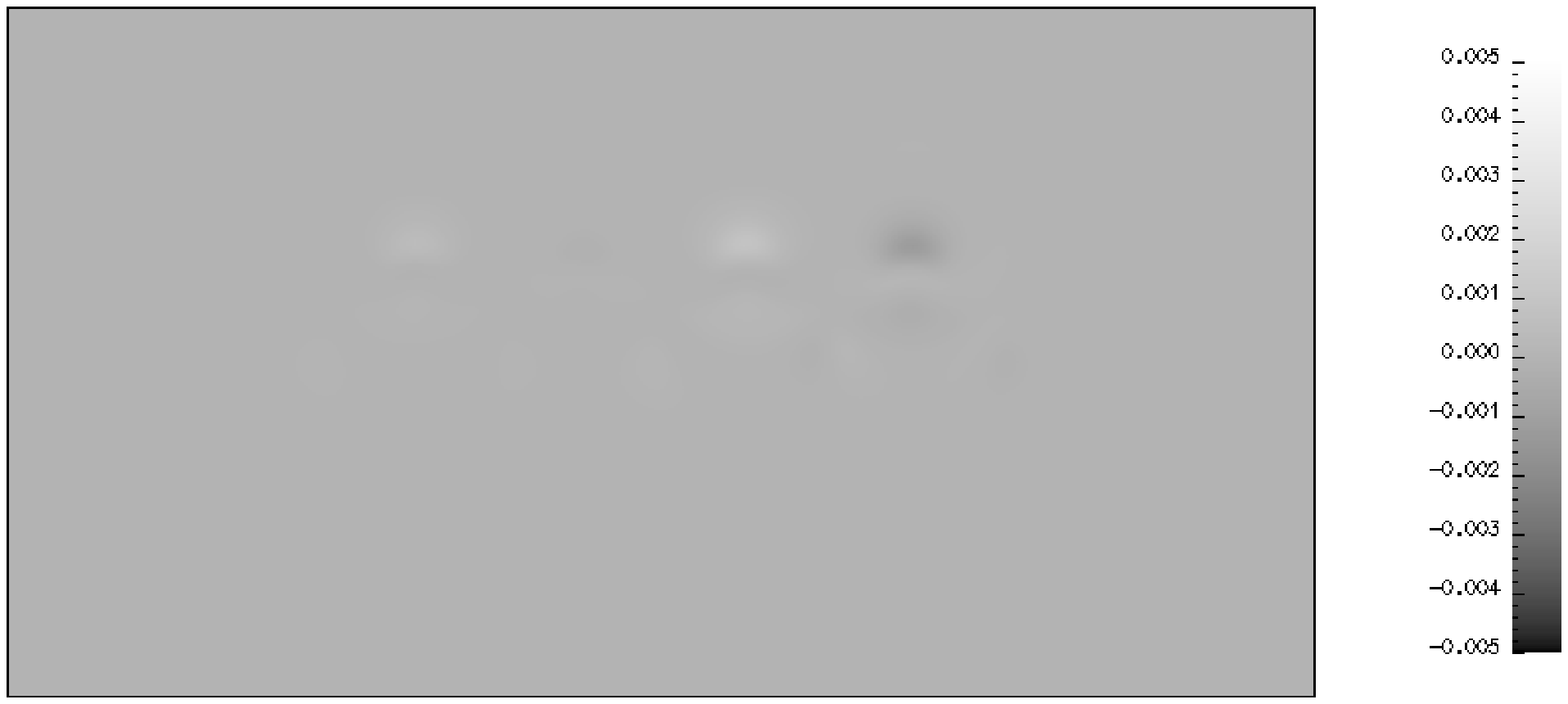}
\includegraphics[angle=-90, width=0.49\columnwidth, clip]{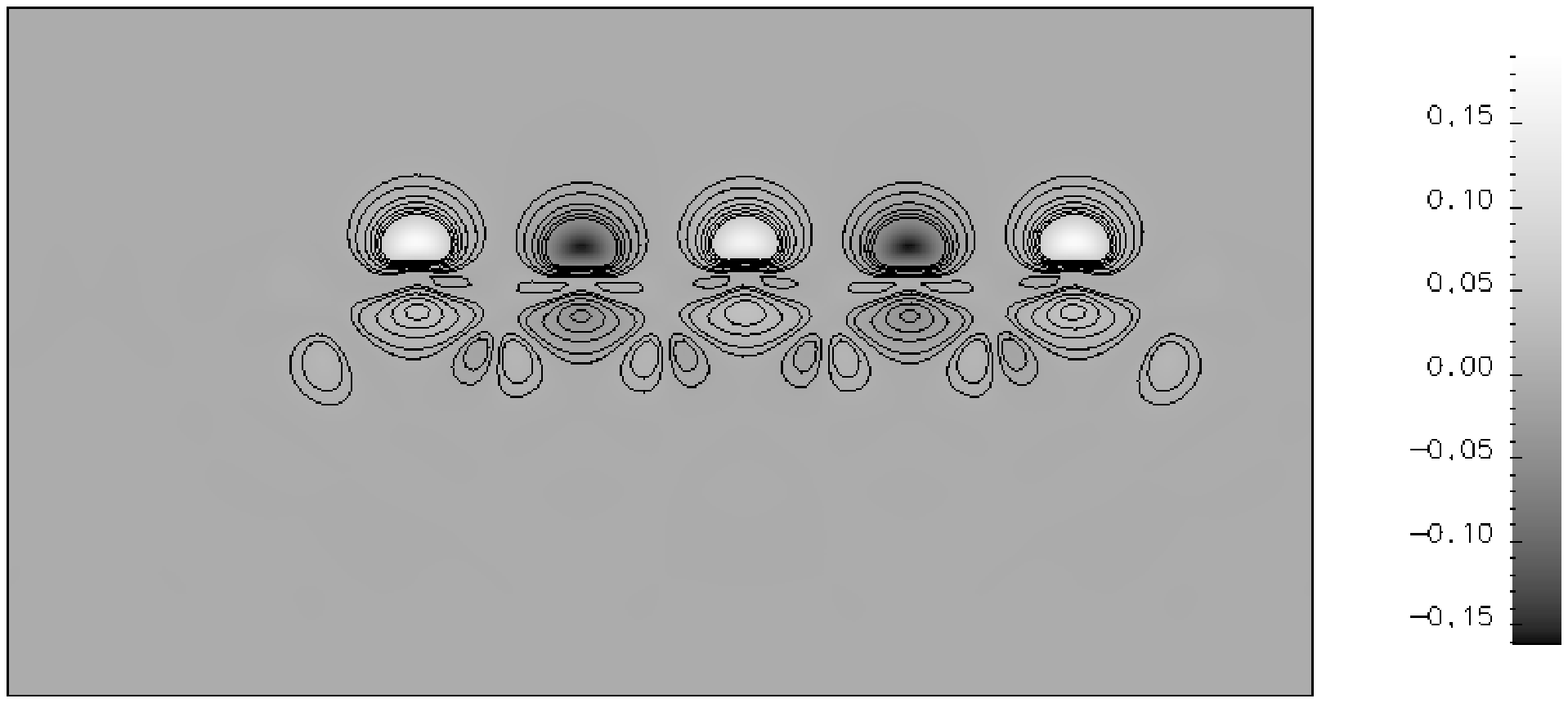}
  \caption{\label{fig:spindiff}Plots of spin difference density for wires of length two,
  three, four and five.  The net spin for both length two and four is
  zero, while for three and five it is one. Note the different scales for odd and
  even length wires.}
\end{figure}

Plots of spin difference for wires of length two, three, four and five
are shown in Fig.~\ref{fig:spindiff}. As expected the results of spin-polarised
calculations for odd-length wires show a strong spin Peierls effect with
alternation between adjacent atoms. However there is only a negligible 
spin Peierls effect in even-length wire, suggesting that the energetic
balance between spin difference and lattice distortion is delicate.
We note that if the net spin zero constraint is removed, an up-up spin
state totalling just more than 1 appears in length 2 and length 4,
that are respectively just 7 and 90 meV more stable.

\section{Conclusions}

In this paper we have presented the results of DFT-GGA simulations of
finite length dangling bond wires on the Si(001)-(2x1) surface.  We
have found that for odd length wires, a spin polarised state with
relatively small displacements ($\sim 0.13$\,\AA) of the atoms is most stable. 
These displacements are in good agreement with those reported from
STM observations\cite{Hitosugi1999} that are in the order of 0.18\,\AA.
These results contrast with both our own non-spin polarised simulations
and those reported previously\cite{Hitosugi1999} that give displacements 
in the order of 0.5\,\AA.

In contrast, for even length (and infinite periodic) wires, a non-spin polarised
state is the most stable. The relative displacement of wire atoms increases strongly
as a function of wire length, from 0.02\,\AA\ for length 2 to $\sim 0.6$\,\AA\  for
length 4 to $\sim 0.8$\,\AA\ for an infinite wire. This is in good qualitative
agreement with STM observations\cite{Hitosugi1999} which range from 0.18\,\AA\ 
for short wires to $\sim 0.6$\,\AA\ in a 13 atom long wire. However, the difference
between experimental and simulated displacements is marked for length 4 wires.

We speculate that this disparity between STM measurements and simulations results
from averaging of a number of states in STM measurements.
As well as the perfectly ordered UDUD and DUDU wires, we have shown\cite{Bird2003}
that solition-containing states are situated at energies very close to the ground state.
Dynamical processes may result in a mixture of these being measured by the STM
process. Further dynamic simulation of the system and of the interaction between 
the surface and STM tip may clarify the situation. 

We conclude that accurate modelling of finite length dangling-bond wires requires
the use of spin-polarised methods for odd length systems and non-spin polarised
approaches are suitable for even length systems. We speculate that, given 
appropriate conditions, spin-polarisation in such systems might be observable
by spin-polarised STM\cite{bode98a}.

\bibliography{DBWireLetter}

\begin{thebibliography}{10}
\expandafter\ifx\csname url\endcsname\relax
  \def\url#1{\texttt{#1}}\fi
\expandafter\ifx\csname urlprefix\endcsname\relax\def\urlprefix{URL }\fi

\bibitem{Shen1995}
T.-C. Shen, C.~Wang, G.~Abeln, J.R.Tucker, J.~Lyding, P.~Avouris, R.~Walkup,
  Science 253 (1995) 1590.

\bibitem{Hitosugi1999}
T.~Hitosugi, S.~Heike, T.~Onogi, T.~Hashizume, S.~Watanabe, Z.~Q. Li, K.~Ohno,
  Y.~Kawazoe, T.~Hasegawa, K.~Kitazawa, Phys.\ Rev.\ Lett. 82~(20) (1999)
  4034--4037.

\bibitem{peierls1955}
R.~Peierls, Quantum Theory of Solids, Clarendon Press, Oxford, 1955, p. 110.

\bibitem{Bowler2001}
D.R.Bowler, A.J.Fisher, Phys.\ Rev.\ B  (2001) 035310.

\bibitem{Jahn1937}
H.A.Jahn, E.Teller, Proc.\ Roy.\ Soc.\ A 161 (1937) 220.

\bibitem{Cho2002}
J.-H. Cho, L.Kleinman, Phys.\ Rev.\ B 66 (2002) 235405.

\bibitem{todorovic02}
M.~Todorovic, A.~Fisher, D.~Bowler, J.\ Phys:\ Cond.\ Mat. 14 (2002) L749.

\bibitem{Bird2003}
C.F.Bird, A.J.Fisher, D.R.Bowler, Submitted to Phys.\ Rev.\ B Cond-mat/0301594.

\bibitem{Watanabe1997}
S.Watanabe, Y.A.Ono, T.Hashizume, Y.Wada, Surf.\ Sci. 386 (1997) 340--342.

\bibitem{pw91}
J.~Perdew, J.~A. Chevary, S.~H. Vosko, K.~A. Jackson, M.~R. Pederson,
  C.~Fiolhais, Phys.\ Rev.\ B\ 46 (1992) 6671.

\bibitem{Kresse1996}
G.Kresse, J.Furthm{\"u}ller, Comp.\ Mat.\ Sci. 6 (1996) 15.

\bibitem{monkhorst76a}
H.~Monkhorst, J.~Pack, Phys.\ Rev.\ B\ 13 (1976) 5188.

\bibitem{Hofer2001}
W.A.Hofer, A.J.Fisher, G.P.Lopinski, R.A.Wolkow, Phys.\ Rev.\ B\ 63 (2001)
  085314.

\bibitem{bode98a}
M.~Bode, M.~Getzlaff, R.~Wiesendanger, Phys.\ Rev.\ Lett.\ 81~(19) (1988) 4256.

\end{thebibliography}
\end{document}